\documentclass[aps,prb,showpacs]{revtex4}

\usepackage{graphicx}
\usepackage{amsmath}
\usepackage{amssymb}
\usepackage{amsthm}

\graphicspath{{./IMAGES/}}

\newcommand{\bra}[1]{\langle #1|}
\newcommand{\ket}[1]{|#1\rangle}

\begin{document}
\title{Nanohelices as superlattices: Bloch oscillations and electric dipole transitions}

\author{C. A. Downing}
\email[]{downing@ipcms.unistra.fr}
\affiliation{Institut de Physique et Chimie des Mat\'{e}riaux de Strasbourg, Universit\'{e} de Strasbourg, CNRS UMR 7504, F-67034 Strasbourg, France}

\author{M. G. Robinson}
\affiliation{School of Physics, University of Exeter, Stocker Road, Exeter EX4 4QL, United Kingdom}

\author{M. E. Portnoi}
\email[]{m.e.portnoi@exeter.ac.uk} \affiliation{School of Physics, University of Exeter, Stocker Road, Exeter EX4 4QL, United Kingdom}
\affiliation{International Institute of Physics, Universidade Federal do Rio Grande do Norte, Natal - RN, Brazil}

\date{\today}

\begin{abstract}
Subjecting a nanohelix to a transverse electric field gives rise to superlattice behavior with tunable electronic properties. We theoretically investigate such a system and find Bloch oscillations and negative differential conductance when a longitudinal electric field (along the nanohelix axis) is also applied. Furthermore, we study dipole transitions across the transverse-electric-field-induced energy gap, which can be tuned to the eulogized terahertz frequency range by experimentally attainable external fields. We also reveal a photogalvanic effect by shining circularly polarized light onto our helical quantum wire. 

\end{abstract}

\maketitle

\section{\label{intro}Introduction}

Helices have played a recurring role in technology, from their use in Archimedes' screws, employed to transfer water in the 3rd century BC, to the modern day usage of giant coils to support buildings against the vibrations of earthquakes. As in many cases, nature arrived at helices before man,\cite{Forterre2011} and examples can be found in plant tendrils, seed pods and seashells at the macroscopic level down to the celebrated double-helix structure\cite{Watson1953} of DNA at the nanoscale.

In condensed matter physics, a series of pioneering works has seen the realization of nanohelices by several different growth and fabrication techniques.\cite{Motojima1990, Amelinckx1994, Zhang1994, Prinz2000, Kong2003, Zhang2003, Zhang2004, Yang2005, Gao2005, Gao2014} The remarkable progress in quality is demonstrated by the recent report of the quantum Hall effect in this novel geometry.\cite{Vorobyova2015} Already there is progress in potential applications from stretchable electronics\cite{Xu2011} to sensing\cite{Hwang2013} to energy storage.\cite{Gao2006} Even DNA itself has been shown to be promising for molecular electronics.\cite{Malyshev2007, Klotsa2005} Recently, a method has been proposed to form a carbon nanohelix by hydrogen doping a graphene nanoribbon,\cite{Zhang2014} which opens up a possible new route to further exploit the superlattice properties of rolled graphene.

It has been shown that the helical motion of electrons subjected to a transverse electric field in chiral carbon nanotubes\cite{Kibis2005a} can give rise to superlattice properties such as Bragg scattering, a precursor to Bloch oscillations.\cite{Bloch1929} Whilst Bloch oscillations have been observed in semiconductor superlattices,\cite{Waschke1993, Feldmann1992} cold atoms in optical lattices,\cite{Dahan1996, Ferrari2006} and photonic structures\cite{Pertsch1999, Morandotti1999, Sapienza2003, Trompeter2006} this phenomenon, as well as negative differential conductance (NDC), have not yet been seen in nanohelices. Importantly from an applications point of view, superlattice physics is the basis for microwave generation by devices such as the quantum cascade laser.\cite{Kazarinov1971, Faist1994}

In this work we consider an electron moving along a semiconductor nanohelix\cite{Kibis2005b, Kibis2007, Kibis2008} in a transverse electric field, which gives rise to a periodic potential and so physics typical of superlattices. Importantly, our presented system has the significant advantage of tunability compared to usual semiconductor superlattices, which is due to the freedom to modulate the applied electric field.  When a longitudinal electric field is applied to the system, we find Bloch oscillations at terahertz (THz) frequencies. The so-called THz gap, the as yet under-utilized part of the electromagnetic spectrum in between microwave and infrared radiation, is increasingly being targeted by device physicists.\cite{Ferguson2002, Lee2007, Hartmann2014} Several THz applications of carbon nanotubes have been proposed,\cite{Portnoi2008, Kibis2007b, Rosenau2009, Batrakov2010, Portnoi2009} and similar useful behavior from nanohelices should also be possible. 

We also consider electric dipole transitions across the energy gap opened up by the transverse-electric-field, which again can be modulated into the THz range, and discuss the optical selection rules and the effects of shining both linearly and circularly polarized light. In particular, we show that a photogalvanic effect arises from shining circularly polarized light along the axis of the helix. This proposal joins a small number of other schemes for producing current from light put forward for curved quantum wires.\cite{Magarill2003, Pershin2005, Entin2009, Entin2010}  

\begin{figure}[hbtp]
 \includegraphics[width=0.4\textwidth]{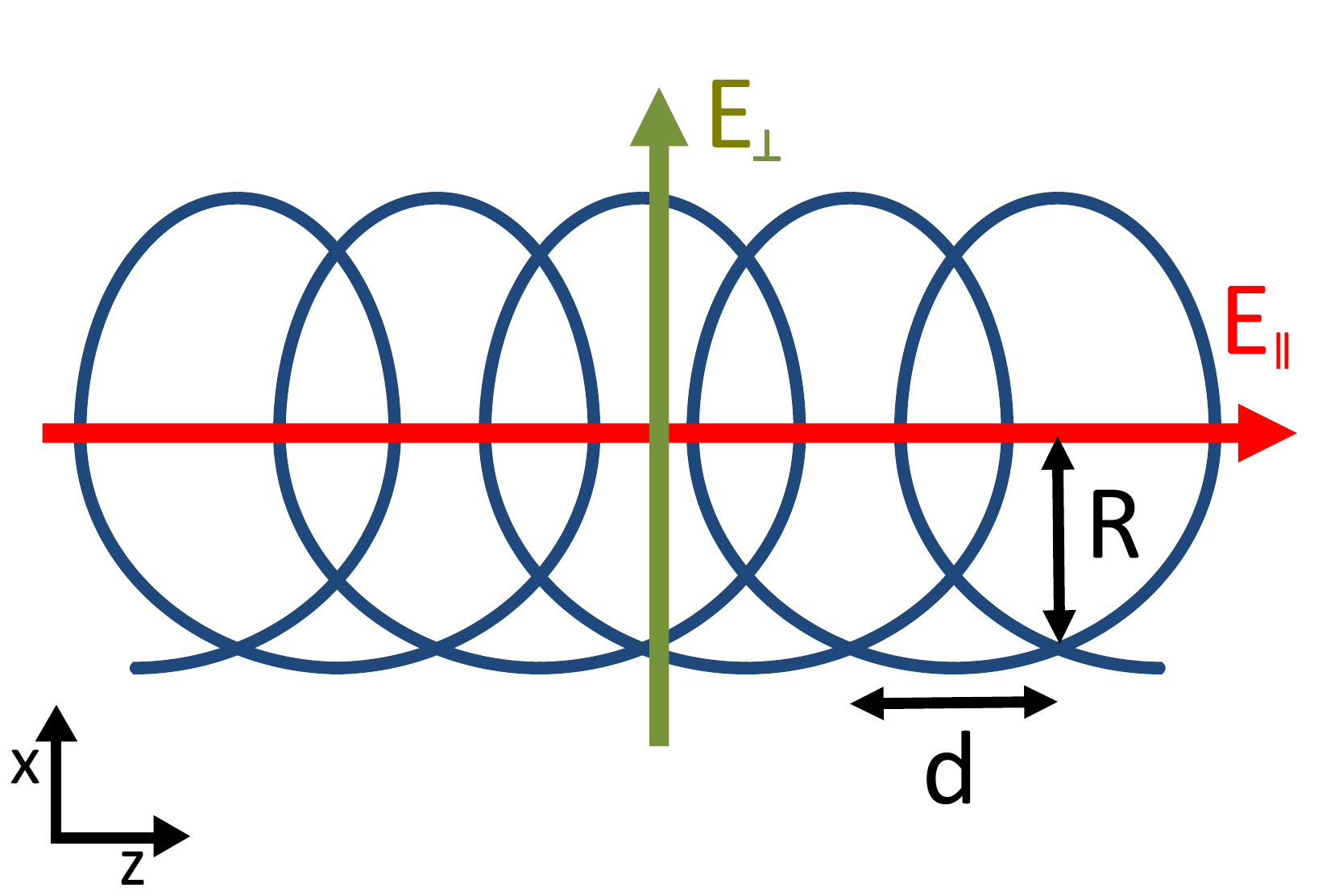}
 \caption{Sketch showing the geometry of the helix considered and the orientation of the applied electric fields. The transverse field $E_{\perp}$ is in place throughout this work, and a further parallel filed $E_{\parallel}$ is utilized in Sec.~\ref{speed}.}
 \label{fig1}
\end{figure}

Since 1900 and the work of Drude\cite{Drude1892}, one-particle models of electrons constrained upon a helical path have been intermittently investigated, principally to study optical activity. Previous works have focused on free motion on helix,\cite{Moore1964, Moore1980} as well as motion in an effective harmonic potential\cite{Maki1996} and with an external static magnetic field.\cite{Wagniere2009} Here we start by studying the case of a transverse electric field, where the Schr\"{o}dinger equation for an electron in a helix is
\begin{equation}
\label{eq11}
	- \frac{\hbar^2}{2 M^{\ast}} \frac{d^2}{d z^2} \psi + e E_{\perp} R \cos \left( \frac{2 \pi z}{d} \right) \psi = \varepsilon \psi,
\end{equation}
where the circular helix is of radius $R$ and pitch $d$, as sketched in Fig.~\ref{fig1}. The effective electron mass $M_e$ is geometrically renormalized to $M^{\ast} = M_e (1+R^2/\bar{d}^2)$, where $\bar{d} = d / 2 \pi $. We have used helical coordinates $\mathbf{r} = \left(x, y, z \right) = \left( R \cos(\xi z / \bar{d}), R \sin(\xi z / \bar{d}), z \right)$, where $\xi = \pm 1$ denotes a left-handed or right-handed helix respectively. It is worth mentioning that in Ref.~\onlinecite{Kibis2008} the coordinate along the helical line was used, but here we have chosen the co-ordinate $z$ (along the axis of the helix) as it is more convenient to study the effects of external fields, and throughout this work we take $\xi = 1$. The one-dimensional periodic potential $V(z) = V(z+ n d)$, which gives rise to superlattice effects, has a period $2 \pi \bar{d}$. In the limit $\bar{d} \to 0$ we recover the standard particle on a ring. Most notably, our electron on a helix system is equivalent to an electron on a quantum ring pierced by a magnetic flux and subject to a lateral electric field.\cite{Fischer2009, Alexeev2012, Alexeev2013, Koshelev2015} In the Aharonov-Bohm ring problem, the role of the quasimomentum in the nanohelices is formally played by a magnetic flux in the units of the flux quantum. Compared to the rings, the helix geometry has the advantage of not needing high magnetic fields to reveal similar physics.

The rest of this work is organized as follows. We study in detail the solution of Eq.~\eqref{eq11} in Sec.~\ref{model} and go on to discuss superlattices properties and Bloch oscillations of nanohelices in Sec.~\ref{speed}. In Sec.~\ref{elements} we investigate electric dipole transitions, whilst we draw some conclusions in Sec.~\ref{conc}. Finally, Appendix~\ref{appendAA} provides the wave equation solution using special functions, Appendix~\ref{appendA} displays the full details of the small matrix results utilized in the main part of the text, and Appendix~\ref{appendDD} considers helices with an inhomogeneous radius.

\section{\label{model}Solution as an infinite matrix}

The exact solution of Eq.~\eqref{eq11} can be expressed in terms of Mathieu functions, as described in Appendix~\ref{appendAA}, where we also provide an analytic expression for the energy spectra with hard-wall boundary conditions. However for our purposes here, it is more illuminating to seek a solution in terms of the Bloch functions
\begin{equation}
\label{eq21}
	\Psi (z) = e^{i k_z z} \sum_m b_m e^{i m z/\bar{d}},
\end{equation}
such that we arrive at an infinite system of equations for coefficients $b_m$
\begin{equation}
\label{eq22}
  [ (q+m)^2 - \lambda ] b_m + u ( b_{m-1} + b_{m+1} ) = 0, \quad m=0, \pm 1, \pm 2, ...
\end{equation}
where we have transformed into the dimensionless quantities $q = k_z \bar{d}$, $\lambda = \varepsilon/ \varepsilon_0(\bar{d}) $, $u = e E_{\perp} R/ 2 \varepsilon_0(\bar{d}) $ and the energy scale is $\varepsilon_0(\bar{d}) = \hbar^2 / 2 M^{\ast} \bar{d}^2 $. Eq.~\eqref{eq22} is equivalent to the $N$-times-$N$ tridiagonal matrix Hamiltonian $H_N$
\begin{equation}
\label{large1}
H_N =
\begin{pmatrix}
 (q+\tilde{N})^2 & u & 0 & \dots \\
  u  & (q+\tilde{N}-1)^2 & u & \dots \\
   0 & u & (q+\tilde{N}-2)^2 & \dots \\
    \vdots & \vdots & \vdots & \ddots \\
\end{pmatrix}
	, 
\end{equation}
where we use the floor function to define $\tilde{N} = \lfloor N/2 \rfloor$. It is noticeable that the Hamiltonian is periodic in the (dimensionless) electron momentum along the helical axis $q$. This periodicity is an important property which is lost when considering small matrices.\cite{Kibis2007, Kibis2008} Additionally, it allows us to restrict our considerations to the first Brillouin zone $-1/2 < q < 1/2$ only, without any loss of physical insight. In fact, for practical calculations it is sufficient to consider the even shorter interval $0 < q < 1/2$ due to the symmetry with respect to the change of sign of $q$.

One needs to find the roots of the resultant characteristic equation formed from Eq.~\eqref{large1} to find the energy bands. An elegant method to find the determinant of our tridiagonal $N$-by-$N$ matrix is via the continuant $K_N$ formalism,\cite{Muir} which observes the recurrence relation
\begin{equation}
\label{eq23}
  K_N = \left[ (\kappa+N-1)^2-\lambda \right] K_{N-1} - u^2 K_{N-2}, 
\end{equation}
subject to the initial conditions $K_0 = 1$, $K_1 = \kappa^2-\lambda$. Carrying out the continuant calculations Eq.~\eqref{eq23} in the dummy variable $\kappa$, the desired determinant of Eq.~\eqref{large1} is then found upon making the replacement $\kappa \to q + 1 - N'$, such that the continuant $K_N  \to \det{(H_N-\lambda I)}$. Here we used the notation $N' = \lceil N/2 \rceil$, with the help of the ceiling function. Then the electron velocity can then be neatly calculated via Euler's chain rule
\begin{equation}
\label{eq25}
  v_z = \frac{1}{\hbar} \frac{\partial \varepsilon}{\partial k_z} = - \frac{\hbar}{2 M^{\ast} \bar{d}} \frac{ \frac{\partial }{\partial q} \det{(H_N-\lambda I)}  }{ \frac{\partial }{\partial \lambda} \det{(H_N-\lambda I)} },
\end{equation}
which, as it maintains periodicity in $q$, is important for physical properties such as Bloch oscillations as we shall see later on in Sec.~\ref{speed}.

\begin{figure*}[htbp]
 \includegraphics[width=0.8\textwidth]{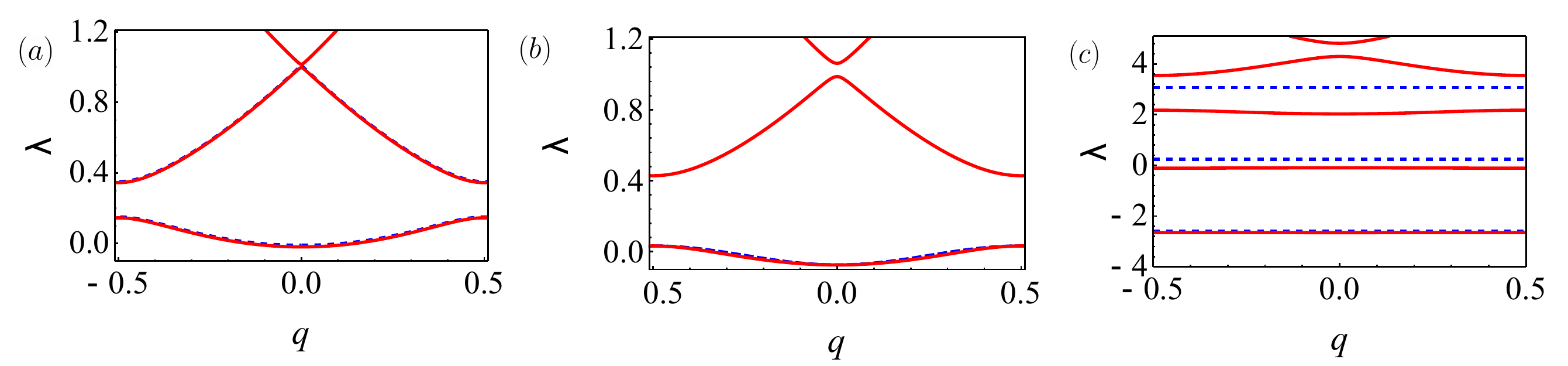}
 \caption{(Color online) Energy spectra for an electron on a helix in a transverse electric field $E_{\perp}$, calculated via Eq.~\eqref{eq22} (solid red lines) and various approximations (dashed blue lines). We present a) the low field regime, here we take $u=0.1$ and also show the 2 band analytical result of Sec.~\ref{2band}; b) the medium field regime, here we take $u=0.2$ and also show the analytic approximation result of Eq.~\eqref{eq052} for the lowest band; c) the high field limit, here we take $u=2$ and also show the flat band mapping results of Eq.~\eqref{eq41}.}
 \label{fig2}
\end{figure*}

We plot in Fig.~\ref{fig2} the lowest energy bands $\lambda$ in three regimes of interest. One can gain further insight into the nature of the energy bands by examining the limiting cases of low and high fields, as in Fig.~\ref{fig2}~(a) and (c) respectively, which can be treated analytically. The intermediate regime of Fig.~\ref{fig2}~(b) can be probed with an expedient analytic replacement.

\subsection{\label{tiny}The low field limit}

In the limit of a weak electric field $u \lesssim 0.2$, one can approximate the lowest energy bands with a truncated matrix $H_N$. The largest polynomial that can always be solved algebraically is quartic, as proved by Abel's Impossibility Theorem. Therefore, one can treat analytically the $N = 2, 3, 4$ band models (please see Appendix~\ref{appendA} for explicit details) by cutting off Eq.~\eqref{large1}. The result can be seen in Fig.~\ref{fig2}~(a) to be in excellent agreement with the result from numerically diagonalizing a large matrix.

\subsection{\label{large}The high field limit}
In this section we consider the limit of a strong perpendicular electric field $u \gg 1$. To our knowledge this limit was not considered analytically for helices, but its manifestations have been clearly seen from numerics for a similar quantum ring problem.\cite{Barticevic2002, Bruno2005} In this regime, the energy of the electron is much smaller than the amplitude of the cosine potential in Eq.~\eqref{eq11} such that the particle is confined near the bottom of the potential where only motion near $z/ \bar{d} = (2 l + 1) \pi$, where $l$ is an integer, needs to be examined. Therefore we treat the periodic potential near its minima as a harmonic oscillator potential
\begin{equation}
\label{eq40}
V_l(z) \to -e E_{\perp} R \left( 1 - \tfrac{1}{2} \xi_l^2 \right), \quad \xi_l = (2l+1)\pi - z/ \bar{d}.
\end{equation}
The required periodicity is accounted for within the nearest neighbor tight-binding method.\cite{Ashcroft} The corresponding Bloch wave functions $\Psi_n$, which serve as the basis for the tight-binding spectrum calculations, are given in terms of properly normalized and centred harmonic oscillator eigenfunctions $\psi_{n}(\xi_l)$. For the lowest subband $n=0$, we have
\begin{equation}
\label{eq44540}
 \Psi_0 = \frac{1}{\sqrt{\mathcal{N}}} \sum_l e^{i \pi q (2 l + 1)} \psi_{0}(\xi_l), \quad \psi_{0}(\xi_l) = \left( \frac{u^{1/2}}{\pi \bar{d}^2} \right)^{1/4} e^{ - \tfrac{u^{1/2}}{2} \xi_l^2 },
\end{equation}
where $\mathcal{N}$ is the number of unit cells, which corresponds to the number of turns of the helix. Carrying out the transfer and overlap integrals for the lowest band with the true cosine potential of Eq.~\eqref{eq11} yields
\begin{equation}
\label{eq489890}
 \lambda_0 = \frac{-2 u e^{- \tfrac{1}{4 u^{1/2}}} + \tfrac{1}{2} u^{1/2} + 2 \gamma_0  \cos(2 \pi q)}{1+2 e^{-\pi^2 u^{1/2}} \cos(2 \pi q)}, 
\end{equation}
\begin{equation*}
\label{eq4898f90}
 \gamma_0 = e^{-\pi^2 u^{1/2}} \left[ u \left( 2 e^{- \tfrac{1}{4 u^{1/2}}} - \pi^2 \right) + \tfrac{1}{2} u^{1/2} \right],
\end{equation*}
which demonstrates the exponential supression of the bandwith with increasing field strength $u$. In the limit of an impenetrable harmonic potential, the mapping to the harmonic oscillator becomes exact and the bandstructure can be written compactly as flat bands
\begin{equation}
\label{eq41}
	\lambda_n = -2u + u^{1/2} (1+2n), \quad n=0,1,2,...
\end{equation}
which for the lowest level $\lambda_0 = u^{1/2} -2u $ has been found to be a good approximation even for $u = 2$ (where the relative error in the first Brillouin zone is $\Delta \lambda_0 / \lambda_0 < 2.5\%$) as shown in Fig.~\ref{fig2}~(b). Therefore, when $u \gg 1$ and further flat bands appear, Eq.~\eqref{eq41} provides the most important term and demonstrates how the system exhibits dispersionless band physics, and consequently a high density of states, in the strong field regime. Notably, systems with flat bands have recently attracted significant attention due to enhanced interaction effects and electronic instabilities which can arise from the high density of states.\cite{Tang2014}

\subsection{\label{period}The periodic approximation}

As was previously mentioned, the periodicity of the band structure cannot be restored via calculations from small truncated matrices. The most convenient way to ensure periodicity for intermediate values of $u$ is to use the following function to describe the lowest band
\begin{equation}
\label{eq052}
	\lambda_0 = \alpha + \tfrac{1}{2} \beta \left[ 1 - \cos(2 \pi q) \right],
\end{equation}
as is familiar from the solution of a one-dimensional periodic potential in the tight-binding approximation. The parameters $\alpha$ and $\beta$ are found from fitting Eq.~\eqref{eq052} to the large matrix result at the center and edge of the first Brillouin zone $q = (0, 1/2)$ respectively. In the limit $u \gg 1$ the parameter $\alpha \to \lambda_0$, as defined in Eq.~\eqref{eq41}. This approximation given by Eq.~\eqref{eq052} is accurate to within 1\% relative to $\varepsilon_0$ when $u \sim 0.2$, as is shown in Fig.~\ref{fig2}~(b). Crucially  this approximation ensures periodicity of the solution which is essential for a proper treatment of Bloch oscillations, which now follows.

\section{\label{speed}Bloch oscillations}

\begin{figure*}[htbp]
 \includegraphics[width=0.8\textwidth]{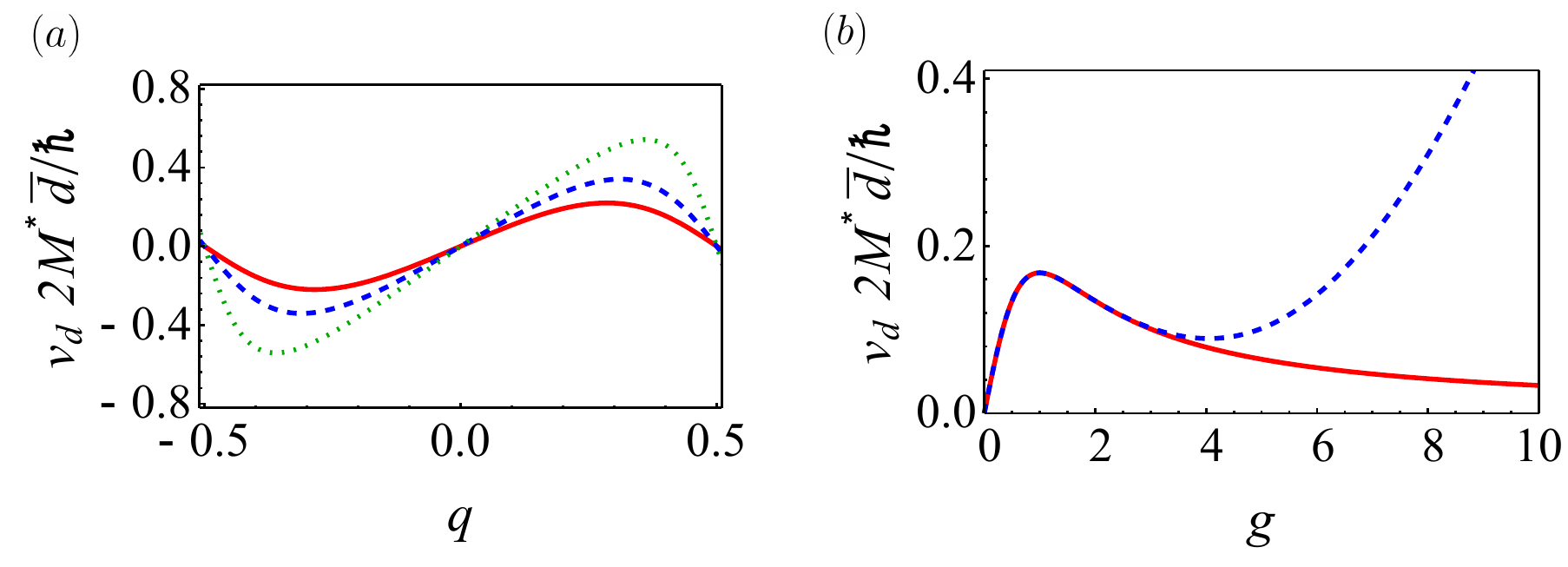}
 \caption{(Color online) (a) Free electron velocity in the first Brillouin zone, with $u=0.3$ (solid red line), $u=0.2$ (dashed blue line) and $u=0.1$ (dotted green line) respectively. (b) Drift velocity as a function of applied field, without (solid red line) and with (dashed blue line) the effect of tunneling from the ground band taken into account. Here, $u=0.2$ and $\varepsilon_0 \tau / \hbar = 10$.}
 \label{fig3}
\end{figure*}

It follows from a consideration of the velocity operator $\hat{v}_z = \tfrac{i}{\hbar} [ \hat{H}, \hat{z} ]$ that the expectation value of electron velocity is
\begin{equation}
\label{eq551}
	\langle v_z \rangle = \frac{\hbar k_z}{M^{\ast}} + \frac{\hbar}{M^{\ast} \bar{d}} \sum_{m} m |b_m|^2,
\end{equation}
which is equivalent to calculations via Eq.~\eqref{eq25}. These expressions describe the free electron velocity and are plotted in Fig.~\ref{fig3}~(a). One notices that the velocity is zero at the center of the Brillouin zone and is suppressed at the edges of the Brillouin zone. A decrease in the (dimensionless) transverse electric field strength $u$ leads to a rescaling of the curve such that the velocity maxima $v_z (q = \tilde{q})$ move further towards the Brillouin zone edges, as governed by
\begin{subequations}
\label{eq5511}
 \begin{gather}
   v_z (\tilde{q}) = \pm \frac{\hbar}{2 M^{\ast} \bar{d}} 2 u f^{3/2},  \label{eq55100} \\
   \tilde{q} = \pm \tfrac{1}{2} \mp u f^{1/2}, \quad f = \left( 2 u \right)^{-2/3} - 1, \label{eq551000} 
 \end{gather}
\end{subequations}
which approaches the limiting velocity $v_z = \pm \hbar / 2 M^{\ast} \bar{d}$ as $u \to 0$. To account for scattering, for example due to phonons or impurities, one may employ the Tsu-Esaki formula\cite{Esaki1970} for drift velocity $v_d = \int_0^{\infty} \exp(-t/ \tau)  \mathrm{d}v_z$, where $\tau$ is the phenomenological scattering time. In an applied field $E_{\parallel}$, and with the semiclassical motion $k_z(t) = e E_{\parallel} t / \hbar$, one obtains in the periodic approximation of Eq.~\eqref{eq052} the following expression for drift velocity in the lowest subband
\begin{equation}
\label{eq552}
	v_d^{0\text{th}} = \frac{\hbar}{2 M^{\ast} \bar{d}} \: \frac{ \pi \beta g}{1+g^2}, \quad g = \frac{E_{\parallel}}{ E_{\tau}},
\end{equation}
with $E_{\tau} = \frac{ \hbar }{ e d \tau}.$ We plot in Fig.~\ref{fig3}~(b) the function Eq.~\eqref{eq552} as a solid red line, showing the maxima at $g = 1$. Beyond this point ($E_{\parallel} > E_{\tau}$) the decreasing drift velocity implies a NDC, since current is proportional to drift velocity via $I = e n v_d$, where $n$ is the electron density. The threshold for observing this NDC effect is an electric field strength of $E_{\parallel} = 1.3 \times 10^3~\text{V/cm}$ and scattering time of $\tau = 0.5~\text{ps}$ for a helix of pitch $d = 10~\text{nm}$.

For higher applied voltages, effects such as Zener tunneling between the first and ground band should be taken into account. Within the WKB approximation, the tunneling probability is an exponential factor\cite{Zener1934} 
\begin{equation}
\label{eq553}
T_{WKB} \approx \exp \left( - \frac{16}{3} \sqrt{2} \pi \: \frac{\varepsilon_0 \tau}{\hbar} \: \frac{u^{3/2}}{g}  \right).
\end{equation}
To calculate the total drift velocity of the 2-band system, one needs to find the drift velocity of the second lowest band $\lambda_1$. This can be done semi-analytically, using the truncated 2-band analysis of Appendix~\ref{2band}. The drift velocity of the first excited band is found to be
\begin{equation}
\label{eq554}
v_d^{1\text{st}}\!= \!\frac{\hbar}{2 M^{\ast} \bar{d}} \left\{ \tfrac{g}{\pi} + e^{-\pi/g} \left[ \gamma (u, g) + \boldsymbol{D}_0 (u, g) - \boldsymbol{D}_2 (u, g) \right]  \right\}
\end{equation}
\begin{equation*}
\label{eq5535}
 \gamma (u, g) = \tfrac{\pi^2 u}{g} \left[ \boldsymbol{H}_1(\tfrac{2\pi u}{g}) - Y_1(\tfrac{2\pi u}{g}) \right] -\tfrac{2 \pi u}{g},
\end{equation*}
where $\boldsymbol{H}_1(\xi)$ is the Struve function and $Y_1(\xi)$ is the Bessel function of the second kind (both of order one) and we introduce the function
\begin{equation}
\label{eq555}
\boldsymbol{D}_n(u, g) = \int_{0}^{1} \frac{ \xi^n e^{\pi \xi / g}}{(\xi^2 +4 u^2)^{\tfrac{n+1}{2}}} \mathrm{d} \xi,
\end{equation}
which can be readily integrated numerically. The total drift velocity of the 2-band system can then be evaluated with the help of Eqs.~\eqref{eq552}-\eqref{eq554} via
\begin{equation}
\label{eq559}
 v_{d} = (1-T) v_{d}^{0\text{th}}+T v_{d}^{1\text{st}},
\end{equation}
and we plot the result in Fig.~\ref{fig3}~(b) as the dashed blue line. This curve reveals that at higher applied fields $E_{\parallel}$, above $g \simeq 4$, the drift velocity (and so current) will again start to increase. Then it follows that the current-voltage characteristic of the nanohelix will be of the so-called `N-type', analogous to Gunn diodes\cite{Gunn1963, Ridley1961} and tunnel diodes.\cite{Esaki1958, Esaki1960} Thus, it is conceivable that nanohelices could be employed in device physics as active elements in amplifiers and generators.

For long electron scattering times, Bloch oscillations at the mini-zone boundary will occur at a terahertz frequency $\omega_B =  e d E_{\parallel} / \hbar = 1.5~\text{THz}$ for nanohelices of pitch $d = 10~\text{nm}$ and $E_{\parallel} = 10^3~\text{V/cm}$, suggesting nanohelices as a useful commodity to resolve outstanding challenges in high frequency generators and amplifiers. We should mention that we do not take into account the effect of charged electric-field domains\cite{Esaki1974b, Buttiker1977, Hyart1997} either stationary or traveling through the superlattice, as in this work we primarily concerned with only a proof of concept of superlattice behavior in nanohelices, however it will be a subject of future research.

\section{\label{elements}Electric dipole transitions}

To understand how our system of a nanohelix subject to a transverse electric field interacts with electromagnetic radiation we calculate the momentum operator matrix element $\mathbf{T} = \bra{a} \mathbf{p} \ket{b}$, which is proportional to the transition dipole moment. Here the momentum operator $\hat{\mathbf{p}}$ is sandwiched between the ground band $\ket{a}$ and second lowest band $\ket{b}$. Explicitly, the self-adjoint momentum operators are\cite{Moore1964}
\begin{subequations}
\label{eq80}
 \begin{gather}
  \hat{\mathbf{p}}_x =  \mathbf{\hat{x}}~i \hbar \tfrac{R/\bar{d}}{1 + R^2/\bar{d}^2} \left[ \sin \left(z/\bar{d}\right) \partial_{z}  + 1 / (2 \bar{d}) \cos \left(z / \bar{d}\right)  \right], \label{tt1} \\
 \hat{\mathbf{p}}_y =  -\mathbf{\hat{y}}~i \hbar \tfrac{R/\bar{d}}{1 + R^2/\bar{d}^2} \left[ \cos \left(z / \bar{d}\right) \partial_{z}  - 1 / (2 \bar{d}) \sin \left(z / \bar{d}\right)  \right], \label{tt2} \\
  \hat{\mathbf{p}}_z = -\mathbf{\hat{z}}~i \hbar \tfrac{1}{1 + R^2/\bar{d}^2} \partial_{z}. \label{tt3} 
 \end{gather}
\end{subequations}
The presence of the trigonometric functions for transverse $(x, y)$ polarized light leads to the optical selection rule that allows transitions  only between states with the angular momentum differing by unity $\left( \Delta m = \pm 1 \right)$, whereas for the $z$-polarized light $\Delta m = 0$.  

\begin{figure}[htbp]
 \includegraphics[width=0.5\textwidth]{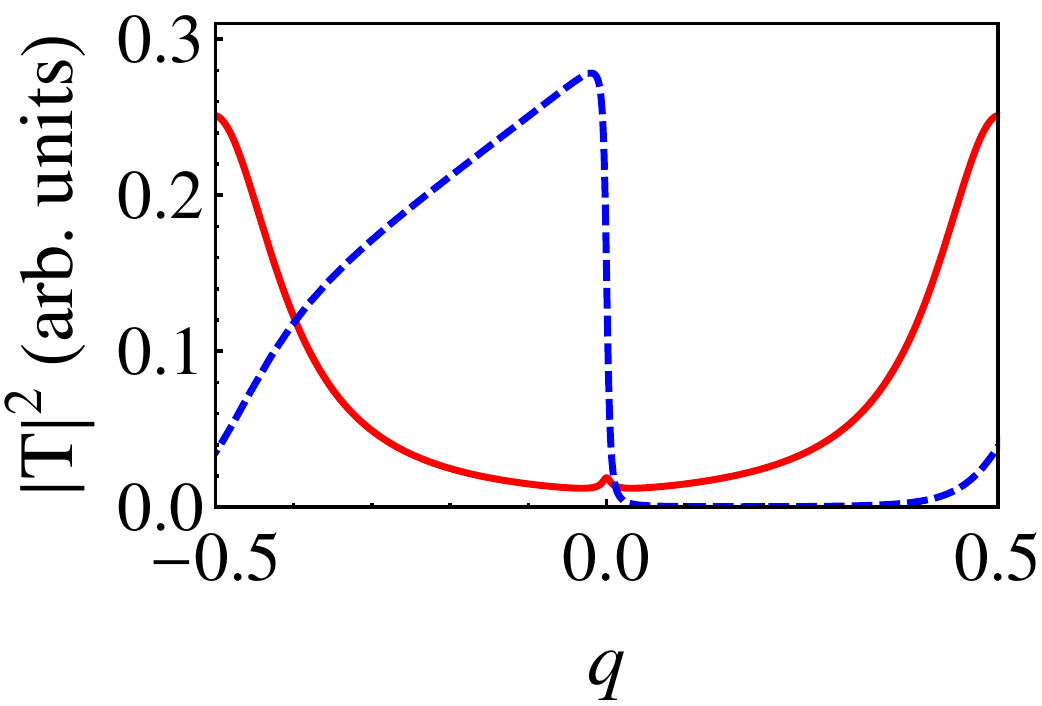}
 \caption{(Color online) Transition dipole moments in the first Brillouin zone as a function of the dimensionless wavevector $q$ of the photoexcited electrons, associated with both (right-handed) circularly polarized light $|T_x + i T_y|^2$ denoted by a dashed blue line and $z$-polarized light $|T_z|^2$ denoted by a solid red line, with $u=0.1$.}
 \label{fig4}
\end{figure}

We present in Fig.~\ref{fig4} the absolute square of the momentum operator matrix element $|T|^2$ for both linearly polarized light normal to the helix axis and for right-handed circularly polarized light propagating along the helix axis. The result for the $z$-polarized light shows a distinctive Mexican-hat-like profile, with global maxima at the edges of the Brillouin zone and a local maximum at the center. This follows from the selection rule: the ground state is almost a pure $m=0$ state, whereas the first excited state is mostly an admixture of the $m= \pm 1$ states, with a contribution from the $m=0$ at the edges of the first Brillouin zone. Thus the peaks at the edges arise from the dominant overlap of the $m=0$ contributions. A small bump near $q = 0$ is due to the second-order in $u$ perturbative corrections, enhanced due to the degeneracy of the $m =\pm 1$ states.

The right-handed circularly polarized light result shows a drastic on-off switching behavior across the two halves of the 1st Brillouin zone, in a left-sided `ski jump' wedge profile with a maximum at the center. The result for left-handed circularly polarized light is simply a mirror image. This result suggests a photogalvanic effect, where one can choose the net direction of charge carriers by shining the appropriate circularly polarized light. Therefore the nanohelix is a promising candidate for a polarization sensitive light detector.

Here we neglected possible inhomogeneities in the helix radius, which are of decreasing importance due to the perpetual increase in nanohelices of high quality. Nevertheless, we discuss inhomogeneities briefly in Appendix~\ref{appendDD} to be quantitative.

\section{\label{conc}Conclusions}

We have investigated two interesting areas of superlattice physics which can arise in nanohelices, a truly tunable superlattice system. Firstly, we showed that the combined effects of a transverse and longitudinal electric field lead to Bloch oscillations in the highly sought-after THz range, and NDC reminiscent of tunneling diodes. Both of these features are attractive for future optoelectronic devices. Secondly, we subjected our system of a helix in a transverse electric field to both circularly polarized electromagnetic waves propagating along the helix axis and the light linearly polarized along the helix axis. We showed that a photogalvanic effect arises in the circularly polarized case.

With the increasingly sophisticated fabrication techniques of complex nanostructures allowing for the assembly of  impressively uniform helices,\cite{Science} we hope our work will inspire experiments on the superlattice and optical properties of nanohelices in the near future and eventually aid the realization of novel THz devices. Future work will inevitably include a study of the influence of a magnetic field on a superlattice behaviour\cite{Alekseev2009} in this helical geometry.

\section*{Acknowledgments}
We would like to thank E. Hendry for fruitful discussions and A. M. Alexeev for a critical reading of the manuscript. We acknowledge financial support from the CNRS and from the ANR under Grant No. ANR-14-CE26-0005 Q-MetaMat, as well as the EU H2020 RISE project CoExAN (Grant No. H2020-644076), EU FP7 ITN NOTEDEV (Grant No. FP7-607521), and the FP7 IRSES projects CANTOR (Grant No. FP7-612285), QOCaN (Grant No. FP7-316432), and InterNoM (Grant No. FP7-612624).

\begin{appendix}

\section{\label{appendAA}Electrons constrained to a nanohelix in a transverse field}

\begin{figure*}[htb]
\centering
 \includegraphics[width=0.6\textwidth]{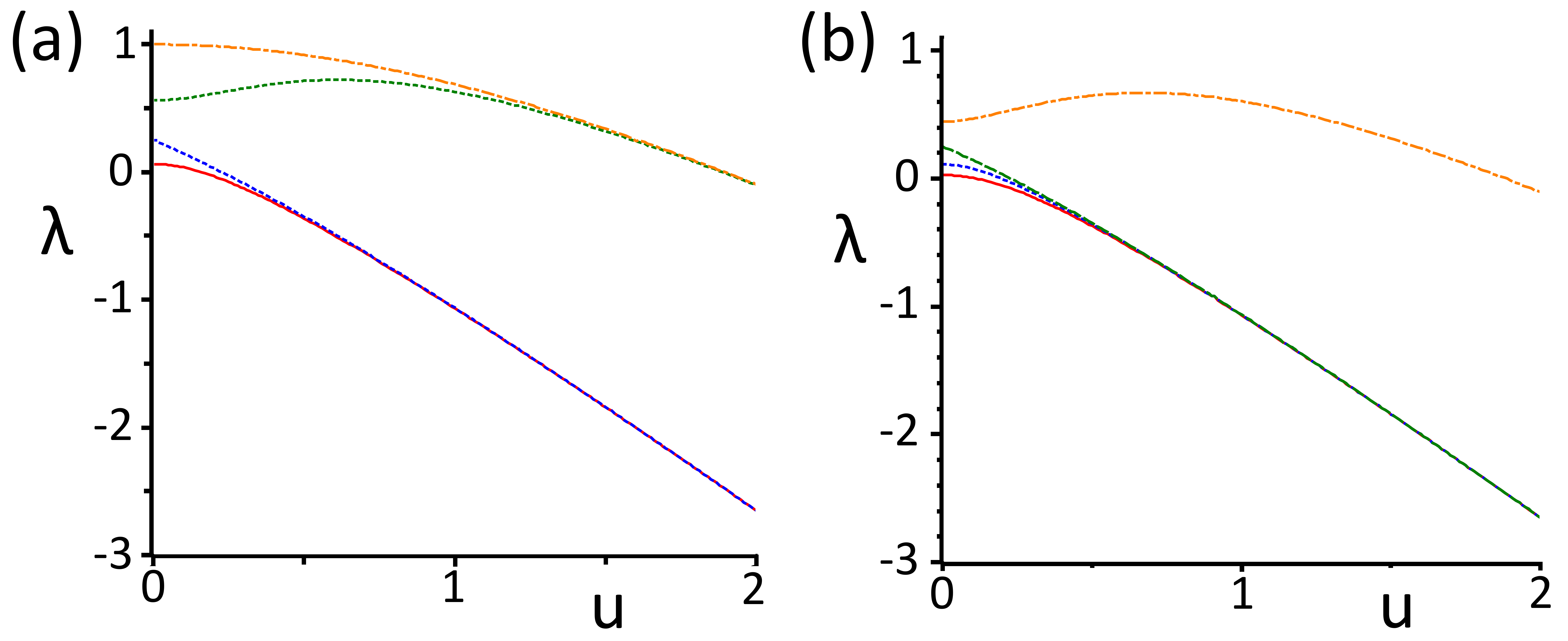}
 \caption{(Color online) Plot of bound state energies for a helix of (a) $\mathcal{N} = 2$ and (b) $\mathcal{N} = 3$ turns as a function of transverse electric field for the four lowest states: the ground state (solid red line) along with the first excited (dotted blue line), second excited (dashed green) and third excited state (dot-dashed orange line).}
 \label{fig5}
\end{figure*}

Eq.~\eqref{eq11} is a limiting case of the more general Lam\'{e} differential equation or Schr\"{o}dinger equation with Jacobi elliptic function potential.\cite{Abramowitz} An extensive study of the Mathieu equation, with an application to an electron moving in a simple cubic lattice, can be found in the classic work of Slater.\cite{Slater} Eq.~\eqref{eq11} has as its two linearly independent solutions the Mathieu sine $S_{4 \lambda}(4 u, z/2\bar{d})$ and Mathieu cosine $C_{4 \lambda}(4 u, z/2\bar{d})$ functions respectively, with characteristic value $4 \lambda$, parameter $4 u$ and variable $z/2\bar{d}$.
 
Here we consider a particle constrained to a nanohelix of $\mathcal{N}$ turns in a transverse field, defined by the potential $V = 2 u \cos (z/\bar{d})$ for $0 \le z/\bar{d} \le 2 \pi \mathcal{N}$ and $V = \infty$ otherwise. With normalization constant $c_n$, the eigenfunctions are
\begin{equation}
\label{eq000000000001}
 \psi_n  = \frac{c_n}{\sqrt{\bar{d}}} S_{4 \lambda}(4 u, z/2\bar{d}), \quad 0 \le z/\bar{d} \le 2 \pi \mathcal{N}
\end{equation}
and $\psi_n = 0$ otherwise. The eigenenergies are given by solutions to the transcendental equation 
\begin{equation}
\label{eq00000001}
 S_{4 \lambda}(4 u, \pi \mathcal{N}) = 0,
\end{equation}
which recovers the infinite square well result in the limit of vanishing potential strength 
\begin{equation}
\label{eq0000000001}
 \lambda_n = \left( \frac{n}{2\mathcal{N}} \right)^2, \quad u \to 0
\end{equation}
where $n$ is an integer. We plot in Fig.~\ref{fig5} the energy spectra of the four lowest-lying states from Eq.~\eqref{eq00000001}. It is noticeable how in progressing from the infinite square well  $u \ll 1$ limit towards the harmonic oscillator $u \gg 1$ limit the neighboring pairs of states coalesce, but intersections are forbidden as is known from the theory of Mathieu functions.

\section{\label{appendA}Solutions of truncated matrices}

\begin{figure*}[tbhp]
 \includegraphics[width=0.8\textwidth]{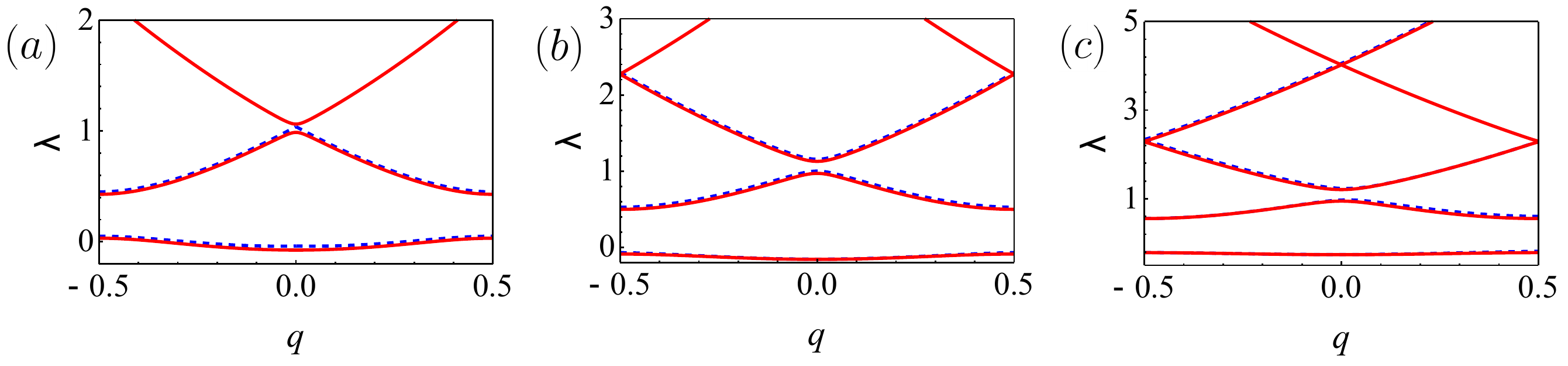}
 \caption{(Color online) Comparison of the analytical results obtained with small matrices (blue dashed lines) with numerical results (solid red lines) from Eq.~\eqref{eq22}. Here, (a), (b) and (c) correspond to the 2-, 3- and 4-band analytical results with $u=0.2, 0.3, 0.4$ respectively.}
 \label{fig6}
\end{figure*}

For completeness, here we present all of the analytical results in the $N = 2, 3, 4$ band approximations of Eq.~\eqref{large1}. The lowest band is denoted $n=0$ and the higher bands are labeled $n=1, 2, 3$ successively. We plot in Fig.~\ref{fig6} these small matrix models.

\subsection{\label{2band}2-band approximation}

The eigenvalues of the two lowest bands $n = 0,1$ are
\begin{equation}
\label{eq001}
 \lambda_{n} = (q+\tfrac{1}{2})^2 + \tfrac{1}{4} \mp s, \quad s = \sqrt{(q+\tfrac{1}{2})^2 + u^2}.
\end{equation}
The accuracy of this truncation can be improved by using Eq.~\eqref{eq001} for $-1/2 < q < 0$ only. Making use of the periodicity of the problem, one may make the substitution $q \to q -1$ in Eq.~\eqref{eq001} and use the resulting expression for $0 < q < 1/2$. One then finds this is a reasonable approximation for both bands for $u \lesssim  0.2$ (the error in $\varepsilon$ in the first Brillouin zone does not exceed 5\%, relative to $\varepsilon_0$). This model also well describes the position and size of the peaks and troughs of the free electron velocity via Eq.~\eqref{eq5511}.

\subsection{\label{3band}3-band approximation}

The lowest three bands $n = 0,1,2$ are arrived at via Cardano's formula with Vieta substitution
\begin{equation}
\label{eq003}
 \lambda_{n} = \tfrac{2}{3} + q^2 + \tfrac{2}{3} \sqrt{1 + 12q^2 + 6u^2} \cos \left( \tfrac{\theta}{3} + \tfrac{\pi}{3} [n-2][3n-1] \right),
\end{equation}
\begin{equation}
\label{eq004}
\text{where}\quad \cos \left( \theta \right) = \frac{36 q^2 - 9u^2 - 1}{\left( 1 + 12q^2 + 6u^2 \right)^{3/2}},
\end{equation}
which is a good approximation for the three lowest bands when $u \lesssim  0.2$ (such that the error in $\varepsilon$ the first Brillouin zone is below 1.5\%, relative to $\varepsilon_0$). This model also tells us that at the edge of the first Brillouin zone the bandgap between the ground and first excited state is $2 u$, a result which is utilized in Eq.~\eqref{eq553}.  

\subsection{\label{4band}4-band approximation}

The formed quartic equation $\lambda^4 + b \lambda^3 + c \lambda^2 + d \lambda + e = 0$, where 
\begin{subequations}
\label{eq310}
 \begin{gather}
  b = -6 - 4q - 4q^2, \label{coupled1} \\
  c = 9 + 8q + 14q^2 + 12q^3 + 6q^4 - 3u^2, \label{coupled2} \\
  d = -4 -4q -2q^2 -10q^4 -12q^5 -4q^6 + 6qu^2 + 6q^2u^2 + 11u^2, \label{coupled3} \\
  e = 4q^2 + 4q^3 -7q^4 - 8q^5 + 2q^6 + 4q^7 + q^8 -8qu^2 -11q^2u^2 - 6q^3u^2 - 3q^4u^2 - 8u^2 + u^4, \label{coupled4} 
  \end{gather}
\end{subequations}
has roots governing the four lowest bands $n = 0,1,2, 3$, given by Ferrari's quartic formula
\begin{subequations}
\label{eq31}
 \begin{gather}
  \lambda_{0, 1} = - \tfrac{1}{4}b -S \mp \tfrac{1}{2} \left( -4S^2 - 2P + \tfrac{U}{S} \right)^{1/2}, \label{coupledone} \\
  \lambda_{2, 3} = - \tfrac{1}{4}b +S \mp \tfrac{1}{2} \left( -4S^2 - 2P - \tfrac{U}{S} \right)^{1/2}, \label{coupledtwo} 
 \end{gather}
\end{subequations}
where
\begin{subequations}
\label{eq32}
 \begin{gather}
  P = c - \tfrac{3}{8} b^2, \quad U = d + \tfrac{1}{8}b^3 - \tfrac{1}{2} bc, \label{one} \\
  S = \tfrac{1}{2} \left( -\tfrac{2}{3}P + \tfrac{1}{3} \left(Q + \tfrac{R}{Q} \right) \right)^{1/2}, \label{two} \\
	Q = \left( \tfrac{1}{2} T + \tfrac{1}{2} \left( T - 4 R^3 \right)^{1/2}  \right)^{1/3},  \label{three} \\
	R = c^2 - 3 b d + 12 e, \label{four} \\
	T = 2 c^3 - 9 bcd + 27b^2e + 27d^2 - 72ce. \label{five}
 \end{gather}
\end{subequations}
The four band approximation is excellent for $u \lesssim  0.2$ (where the error in $\varepsilon$ in the first Brillouin zone is below 1.5 \% for the lowest four lowest bands, relative to $\varepsilon_0$).

Whilst larger truncated matrix models cannot be expressed algebraically, analytical expression can be obtained with the help of special functions for the 5-band (with Jacobi theta functions), 6-band (with Kamp\'{e} de F\'{e}riet functions) and 7-band (with hyperelliptic functions and associated theta functions of genus 3) models. In general, any algebraic equation can be solved with modular functions, and the roots should be expressible with hyperelliptic integrals and high genus theta functions.\cite{King}

\section{\label{appendDD}A 1-D quantum helix with an inhomogeneous radius}

Despite remarkable advances in the synthesis of nanohelices, there is still some degree of inhomogeneity in the radius of a given helix. We consider the effect of a changing helix radius\cite{Exner, Zampetaki} along the helix axis, in the variable radius coordinates 
\begin{equation}
\label{eqd1}
	\mathbf{r} = \left( R (z) \cos( z / \bar{d}), R (z) \sin( z / \bar{d}), z \right).
\end{equation}
The equation of motion equation becomes a free Schr\"{o}dinger equation in the new dependent variable
\begin{equation}
\label{eqd2}
	\xi (z) = \int^z h(z') \mathrm{d}z', \quad h(z) = \left( 1 + R(z)^2/\bar{d}^2 + R'(z)^2 \right)^{1/2}
\end{equation}
where $R'(z)$ represents a derivative with respect to $z$, such that the eigensolutions are 
\begin{equation}
\label{eqd3}
	\psi_n  = \frac{c_n}{\sqrt{\bar{d}}} \sin \left( k \xi \right), \quad \varepsilon_n = \frac{\hbar^2}{2 M} \left( \frac{n \pi}{\xi_{\mathcal{N}}} \right)^2,
\end{equation}
where $k = \left( 2 M \varepsilon / \hbar^2  \right)^{1/2}$ and $\xi_{\mathcal{N}} = \xi (2 \pi \mathcal{N} \bar{d}) $. The limiting case of a homogeneous helix $R(z) = R$ recovers the solution $ \varepsilon^{\text{homo}}_n = \frac{\hbar^2}{8 M^{\ast} \bar{d}^2} \left( \frac{n}{\mathcal{N}} \right)^2$, as found in Eq.~\eqref{eq0000000001} in the limit of zero transverse field. Considering a Gaussian bump inhomogeneity described by $R(z) = R \left( 1 + \gamma \exp(-z^2/\bar{d}^2)\right)$, one finds the relative energy eigenstates do not deviate dramatically from the homogeneous case. For example, $\varepsilon^{\text{bump}}/\varepsilon^{\text{homo}} \approx 0.95$ for a helix of parameters $R/\bar{d} = 1$ and $\mathcal{N} = 1$, with bump parameter $\gamma = 0.3$.

\end{appendix}

\end{document}